\documentclass[journal]{IEEEtran}

\usepackage{placeins}
\usepackage{amsmath}
\usepackage{algorithm}
\usepackage{algorithmic}
\usepackage{booktabs}
\usepackage{multirow}
\usepackage[T1]{fontenc}    % use 8-bit T1 fonts
\usepackage{hyperref}       % hyperlinks
\usepackage{url}            % simple URL typesetting
\usepackage{booktabs}       % professional-quality tables
\usepackage{amsfonts}       % blackboard math symbols
\usepackage{nicefrac}       % compact symbols for 1/2, etc.
\usepackage{microtype}      % microtypography
\usepackage{xcolor}         % colors
\usepackage{graphicx}
\usepackage{float}
\usepackage[numbers,sort&compress]{natbib}
\usepackage{caption} % For better control over table captions
\newfloat{figtab}{htb}{fgtb}
\makeatletter
  \newcommand\figcaption{\def\@captype{figure}\caption}
  \newcommand\tabcaption{\def\@captype{table}\caption}
\makeatother

\def\BibTeX{{\rm B\kern-.05em{\sc i\kern-.025em b}\kern-.08em
    T\kern-.1667em\lower.7ex\hbox{E}\kern-.125emX}}

\title{Online Inference of Human Intention as a Latent Control State from Single-Trial EEG}

\label{sec:title}
\author{
\IEEEauthorblockN{
Xiaowei Jiang\textsuperscript{1}\textsuperscript{*}, 
Daniel Leong\textsuperscript{1}, 
Yu-Cheng Chang\textsuperscript{1}, 
Thomas Do\textsuperscript{1}, 
Chin-Teng Lin, \emph{IEEE Fellow}\textsuperscript{1}\\
}

\IEEEauthorblockA{\textsuperscript{1}Computational Intelligence and Brain-Computer Interfaces lab,\\ Faculty of Engineering and Information Technology, University of Technology Sydney}

\thanks{\textsuperscript{*}Corresponding author: Xiaowei Jiang. Email: xiaowei.jiang-1@student.uts.edu.au}
\thanks{ \textcolor{black}{This work was supported in part by the Australian Research Council (ARC) under discovery grant DP250103612 and DP260101395, ARC Research Hub for Human-Robot Teaming for Sustainable and Resilient Construction (ITRH) grant IH240100016, and Australian National Health and Medical Research Council (NHMRC) Ideas Grant APP2021183. Research was also sponsored in part by the Australia Advanced Strategic Capabilities Accelerator (ASCA) under Contract No. P18-650825 and ASCA EDT DA ID12994, and the Australian Defence Science Technology Group (DSTG) under Agreement No: 12549.}}
}

\begin{document}
\maketitle

\begin{abstract}
Human intention can be modeled as a latent internal state that modulates how sensory information is evaluated and committed to action in human-machine systems.
However, most existing brain-computer interfaces (BCIs) rely on control signals tightly coupled to externally imposed stimulation and do not explicitly infer whether perceived stimuli align with a user’s internal goals.
Here, we investigate whether intention can be inferred as a latent, goal-dependent state from single-trial electroencephalography (EEG).
We introduce a stimulus-based paradigm in which intention is specified by an internally cued target category, while object identity vary independently across stimuli.
To estimate intention under single‑trial neural variability, we propose an interpretable fuzzy prototype‑based network; it offers transparent neural‑pattern alignment by mapping each trial onto interpretable fuzzy prototypes that encode intention‑specific dynamics.
The model encodes intention-related neural dynamics using a compact set of fuzzy prototypes with soft memberships, enabling robust decoding without reliance on engineered mediating stimuli.
Experimental results demonstrate reliable within-subject single-trial intention decoding that outperforms representative deep learning baselines (Accuracy: $93.22\% \pm 3.21\%$), with online validation further confirming real-time feasibility (Accuracy: $70.11\% \pm 10.87\%$).
Together, these findings advance intention-aware BCIs from stimulus-driven detection toward principled inference of goal-dependent internal states.
\end{abstract}


\begin{figure}[t]
    \centering
    \includegraphics[width=1\linewidth]{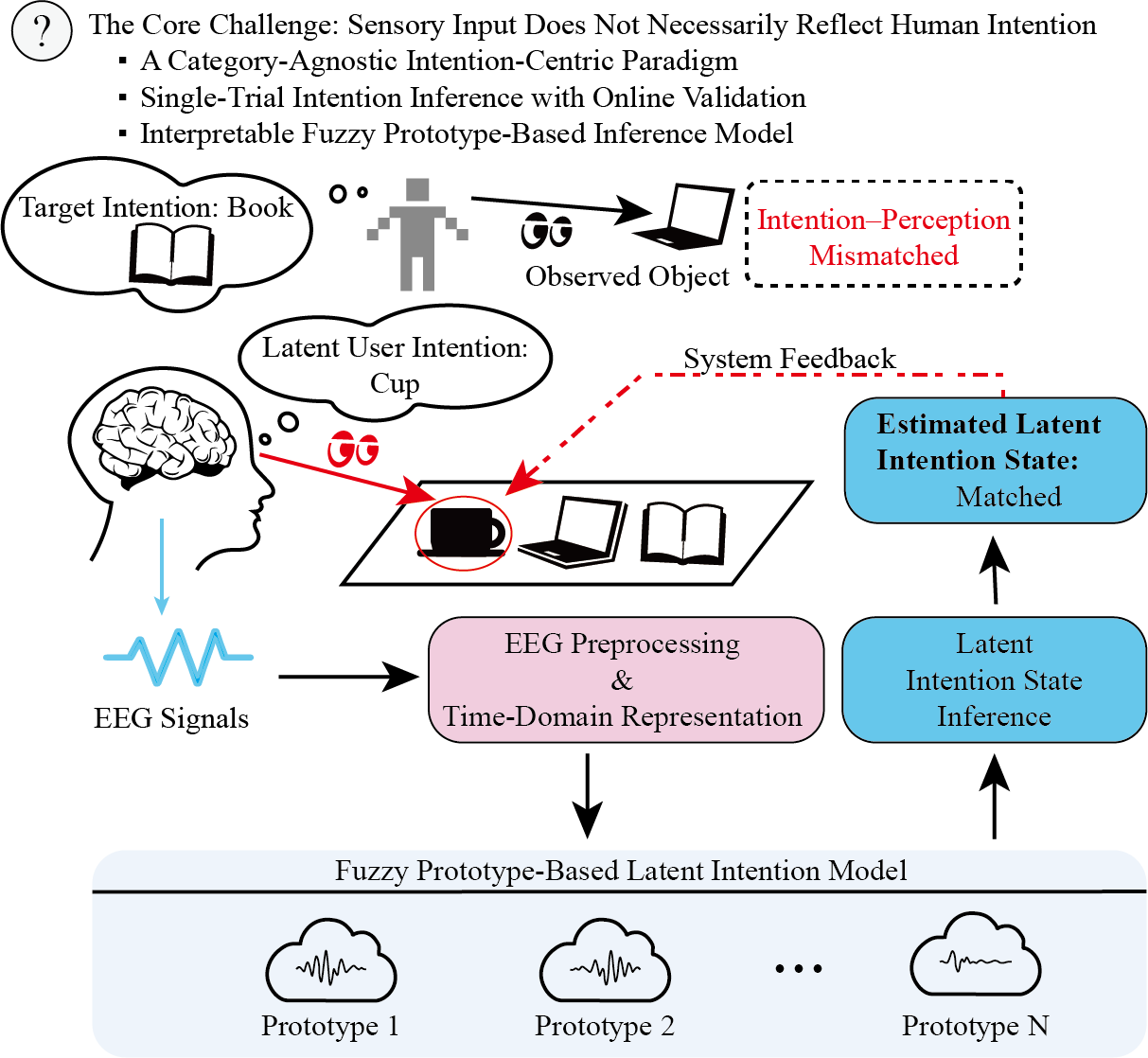}
    \caption{
    \textbf{Intention-centric BCI framework for resolving perception–intention mismatch.}
    The system infers a user’s latent intention state from single-trial EEG using a fuzzy prototype-based model, and provides intention-aware feedback within a closed-loop perception-inference architecture.
    }
    \label{fig:graphic_abstract}
\end{figure}

\section{Introduction}

Human perception is not a passive readout of sensory input, but an active, goal-dependent process in which identical stimuli can be evaluated in fundamentally different ways depending on what the observer intends to achieve~\cite{Memelink2013IntentionalWeighting,Harel2014TaskContext}.
In everyday vision, people constantly encounter objects that are seen but ignored, noticed but rejected, or briefly processed without being selected for action.
Thus, what is physically presented is not necessarily what is behaviorally relevant.
This dissociation highlights a critical distinction between \emph{what is seen} (the externally presented visual object class) and \emph{what is wanted} (the internally evaluated goal relevance of that object), implying the presence of an internal intention state that regulates how sensory evidence is evaluated and committed to action~\cite{caoBuildingEEGbasedCAD2022,trieschWhatYouSee2003}.

Neuroscience studies provide strong support for this view.
When stimulus properties are held constant, neural responses can differ markedly depending on task goals, target definitions, or internal expectations~\cite{Alaidroos2012,Harel2014,Bansal2014}.
These findings indicate that perceptual processing reflects an interaction between external sensory input and latent internal states encoding current goals.
Importantly, such modulation does not eliminate stimulus dependence; rather, it reshapes how stimulus information is interpreted and evaluated.
For brain-computer interfaces (BCIs), this implies that decoding based solely on stimulus-locked activity may fail to capture whether a stimulus is actually relevant to the user’s intention.

From a control and cybernetic perspective, intention can be formalized as a latent internal state that governs perceptual evaluation and perception-action coupling~\cite{powers2024behavior,Daniel1995InternalModel,Friston2011ActionUnderstanding}.
Intention does not correspond to a single observable event, nor is it equivalent to stimulus detection.
Instead, it reflects an evolving internal variable that determines whether incoming sensory information is treated as relevant for the current goal.
An intention-aware BCI must therefore infer this latent state from noisy neural observations, distinguishing between stimuli that are merely perceived and those that are evaluated as goal-relevant, rather than detecting externally defined events alone.

% Contemporary theories of decision-making and conscious access converge on this distinction.
% The global workspace framework proposes that reportable and action-guiding perception arises when sensory evidence is evaluated with respect to task goals, rather than through stimulus presence alone~\cite{Dehaene2011}.
% Behavioral studies further show that attentional and decision resources are preferentially allocated to goal-relevant information, even when multiple stimuli are simultaneously present~\cite{Vogt2010,Dijksterhuis2010}.
% Electrophysiological evidence dissociates early sensory encoding from later evaluative and decision-related stages, demonstrating that late neural activity reflects goal-dependent commitment rather than stimulus encoding per se~\cite{Pitts2014}.
% Together, these findings suggest that intention decoding should target neural signatures of evaluative relevance rather than purely stimulus-driven responses.

\begin{figure*}[t!]
    \centering
    \includegraphics[width=1\linewidth]{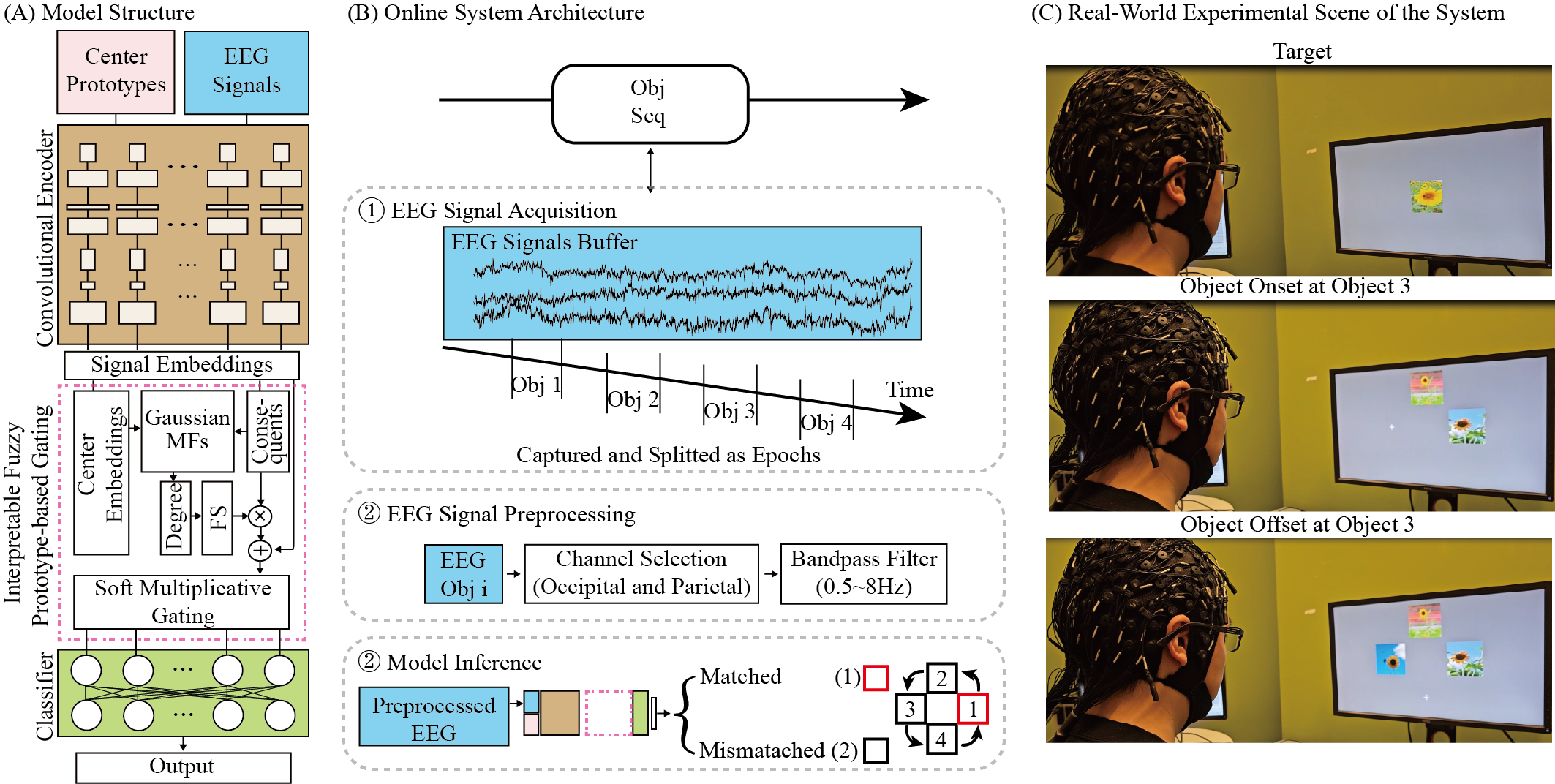}
    \caption{
        Overview of the proposed framework and online system.
        (A) Model structure, consisting of a Convolutional encoder, a fuzzy prototype-based gating modeling module, and a classification head.
        (B) Online system architecture for real-time EEG acquisition, processing, and model inference.
        (C) Real-World Experimental Scene of the System (Object 3 is at the left of the monitor).
    }

    \label{fig:model}
\end{figure*}

A well-established neural correlate of such goal-dependent evaluation is the P3b component, a late parietal event-related potential associated with target detection, context updating, and decision commitment.
Crucially, the P3b does not index simple stimulus matching.
Instead, its amplitude and latency scale with internal target representations~\cite{Hout2015,Rajsic2017}, expectations about relevance~\cite{Brookhuis1983,Kopp2016}, and motivational significance~\cite{Begleiter1983,Kaya2022}.
P3b modulation can be observed when physical stimulation is identical and only task relevance differs, indicating that it reflects an internal evaluative process rather than stimulus-driven encoding.
Evidence from dissociation and no-report paradigms further suggests that the P3b reflects post-perceptual evaluation and decision-related processing rather than conscious experience~\cite{Cohen2020,Dembski2022}, making it particularly suitable as an observable marker of internal decision relevance for cybernetic systems.

Despite decades of progress, most existing BCIs remain tightly coupled to externally imposed stimulation paradigms.
Canonical systems such as P300 spellers and steady-state visual evoked potential (SSVEP) interfaces rely on predefined event structures or rhythmic modulation to elicit discriminable neural responses~\cite{Elshout2009,Wolpaw2002,Li2020,Maslova2023,11205327,cao2025emd,cao2025novel}.
Crucially, these paradigms introduce stimulus-irrelevant mediating mechanisms, such as periodic flicker, frequency tagging, or oddball timing, that are not intrinsic to the visual objects themselves but are engineered to render neural signals classifiable.
As a result, decoding performance often reflects sensitivity to artificial control signals rather than inference of whether a perceived stimulus is internally evaluated as relevant to a user’s goal.
Although effective in constrained settings, such systems conflate stimulus detection with intention inference and impose additional cognitive demands, leading to limited flexibility and mental fatigue~\cite{Wolpaw2010,Lotte2018, tran2024multimodal}.
More fundamentally, many BCI approaches continue to decode externally defined control signals instead of determining whether a stimulus is evaluated as goal-relevant by the user.

Recent object-based BCI studies have begun to distinguish between object recognition and object identification by exploiting differences in functional connectivity patterns and task-specific channel configurations~\cite{leong2023ventral, leong2024distinction}.
While these studies represent important progress toward intention-aware object-based BCIs, intention inference in these frameworks remains inherently tied to object category structure.
Identification is implemented through category-specific decoding pipelines whose complexity scales with the number of object classes, rendering large-scale or open-world deployment impractical.
As a consequence, intention is defined relative to predefined object categories rather than as a category-agnostic evaluative state, and the inferred state remains contingent on what is seen rather than whether the stimulus is internally evaluated as relevant to an intended goal.

A further challenge arises from the intrinsic variability of late-stage evaluative neural responses at the single-trial level.
Although the P3b component exhibits moderate-to-high reliability when averaged across trials~\cite{Cassidy2012,Saville2012}, its single-trial amplitude, latency, and scalp distribution vary substantially even within the same individual.
Neural variability increases after approximately 200~ms post-stimulus~\cite{Maleki2025}, coinciding with the emergence of decision- and evaluation-related activity, and is further modulated by factors such as fatigue and vigilance~\cite{Haubert2018,Liu2024}.
As a result, relying on P3b alone as a decoding feature is often insufficient for robust intention inference under realistic single-trial settings.
Instead, the P3b should be viewed as a reliable biomarker embedded within a broader set of temporally evolving neural signatures, including multiple event-related potential (ERP) components and concomitant spectral dynamics.

Fuzzy learning provides a principled framework for modeling such heterogeneous and uncertain neural representations~\cite{jiang2025fuzzy,jiang2025interpretable,jiang2025ifuzzmeta,lin1996neural, 10183374,106218}.
By operating directly on time-domain EEG signals that jointly encode P3b-dominated ERP structure and distributed frequency information, fuzzy systems represent latent cognitive states through graded memberships rather than hard class assignments.
This formulation naturally accommodates single-trial uncertainty, temporal variability, and overlapping neural responses without assuming stationary or sharply separable feature distributions.
In this context, fuzzy inference is particularly well suited for intention decoding, where internal evaluative states are continuous, uncertain, and only indirectly observable from noisy stimulus-locked EEG, even under within-subject decoding paradigms.

In this work, we investigate whether single-trial EEG can be used to infer human intention when intention is defined as a latent goal state that modulates stimulus evaluation, rather than as a response to externally imposed control signals, as illustrated in Fig.~\ref{fig:graphic_abstract}.
At the beginning of each trial sequence, participants are cued with an abstract goal definition rather than a specific object or object category.
Objects are then presented sequentially at one of four spatial locations, with object identity varying independently of goal relevance.
For each stimulus, participants internally evaluate whether the object matches their current intention.
Thus, although the paradigm is stimulus-driven, the critical variable of interest is whether the same physical stimulus is internally evaluated as goal-relevant or irrelevant.
Under this design, systematic differences in neural activity primarily reflect internal evaluative states rather than stimulus properties or engineered mediating signals.

To infer intention from these observations, we propose a Fuzzy Prototype-Based Latent Intention (Fuzz-PLI) Model for latent state estimation.
The model learns a compact set of spatiotemporal neural prototypes associated with intention-relevant (Match) and intention-irrelevant (Mismatch) evaluations and assigns graded memberships at the single-trial level.
Compared with prior fuzzy EEG decoding frameworks~\cite{jiang2025ifuzzmeta}, the proposed Fuzz-PLI model adopts a TSK-inspired softmax-weighted firing strength~\cite{shihabudheen2018recent}, leading to more stable and discriminative prototype contributions.
This formulation improves decoding robustness under neural variability while preserving transparent rule- and prototype-level interpretability.
Using this framework, we achieve an offline within-subject decoding accuracy of $93.22\% \pm 3.21\%$, and further validate real-time feasibility in an online setting, achieving an average online decoding accuracy of $70.11\% \pm 10.87\%$.

The contributions of this paper are threefold:
\begin{enumerate}
    \item \textbf{Category-agnostic intention inference without mediating stimulation.}
    We introduce a BCI paradigm that infers human intention independently of object category structure and without relying on externally engineered control signals.
    
    \item \textbf{Robust single-trial intention decoding in online settings.}
    We demonstrate that latent intention states can be reliably inferred from single-trial EEG successful real-time online decoding.
    
    \item \textbf{Interpretable fuzzy modeling of evaluative neural dynamics.}
    We propose an improved fuzzy prototype-based network that captures intention-related neural activity and exposes interpretable knowledge through learned prototypes and fuzzy rules.
\end{enumerate}

Together, these results show that even in stimulus-driven environments, human intention can be inferred as a latent evaluative state that determines whether a stimulus is treated as goal-relevant.
This work advances BCI design from detecting what users see toward inferring what users intend the system to act upon, without dependence on stimulus-mediated control mechanisms.

\begin{figure*}[ht!] 
    \centering 
    \includegraphics[width=1\linewidth]{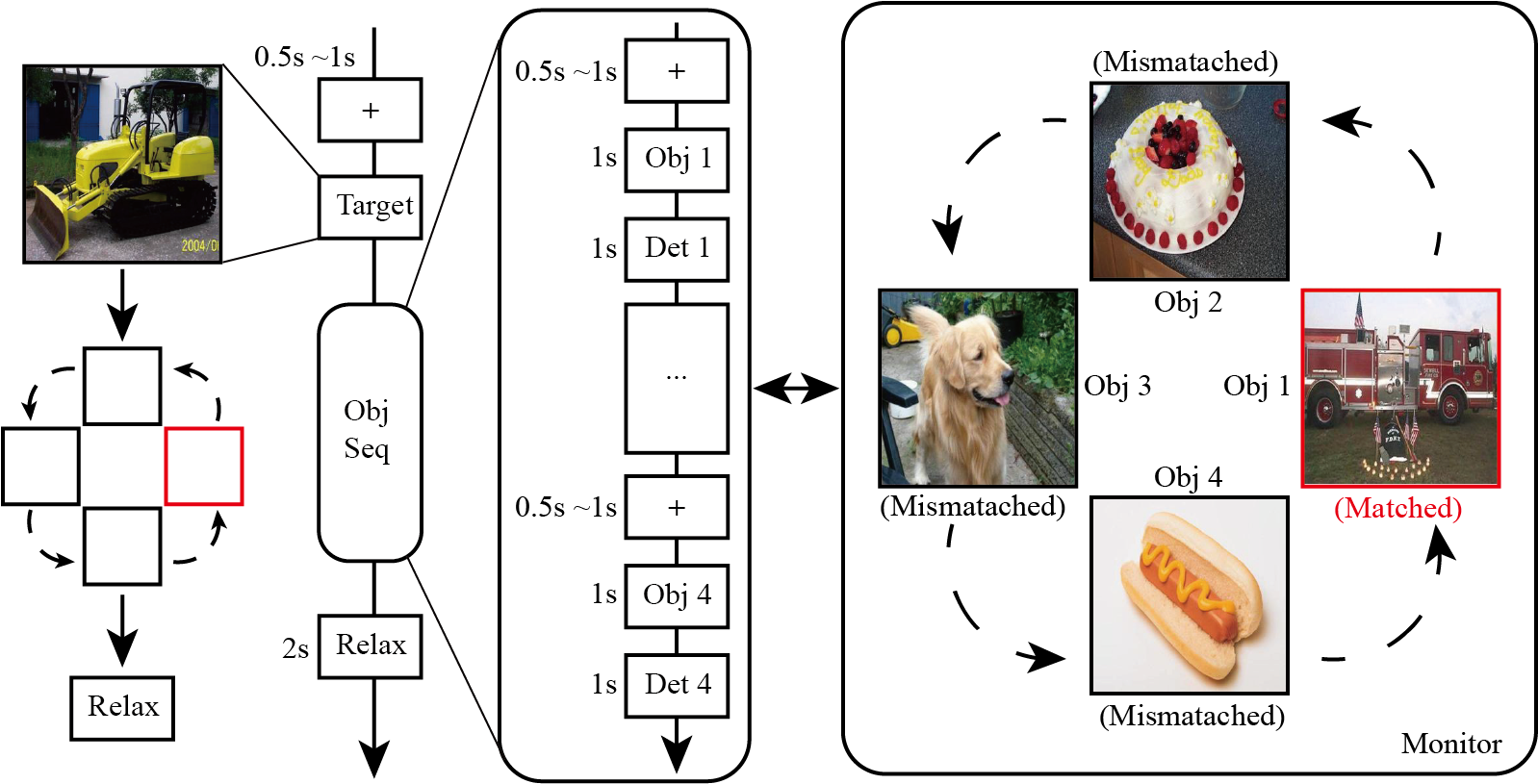} 
    \caption{Category-cued human intention paradigm with offline and online variants. At the start of each sequence, a target category is cued at fixation (Target), followed by a fixation period (+). Four objects (Obj~1\textasciitilde Obj~4) are then presented sequentially at one of four possible locations (up, down, left, right) around fixation. In the offline paradigm, only the object images and fixation are shown, and participants internally judge whether each object matches the cued category. In the online paradigm, each object presentation is followed by a decision interval (Det~1\textasciitilde Det~4) in which the same object is re-displayed with a colored frame indicating the system’s current decision: a black frame denotes a predicted non-target, whereas a red frame denotes a predicted target. Relaxation periods separate successive sequences. The example on the right illustrates a sequence in which only Obj~4 is \emph{intention-relevant} (red frame, Matched), and the remaining objects are \emph{intention-irrelevant} (black frame, Mismatched).} 
    \label{fig:model_exp_design} 
\end{figure*}

\section{Methodology}
\label{sec:methods}

\subsection{Task Definition}

We aim to infer, on a single-trial basis, whether a visually presented stimulus is internally evaluated as \emph{intention-relevant} by the observer.
Intention is defined as a category-agnostic evaluative state reflecting whether the current stimulus is ``what I want'' under the observer's current goal, rather than recognition of a specific object class.
Accordingly, each stimulus presentation is associated with a covert binary goal-relevance condition: \emph{intention-relevant} (matched) and \emph{intention-irrelevant} (mismatched).

For each participant, the EEG is segmented into trial-wise multichannel time series
\begin{equation}
    x_i \in \mathbb{R}^{C \times T}, \quad i = 1,\dots,N_{\text{trials}},
\end{equation}
where $C$ denotes the number of EEG channels and $T$ denotes the number of time samples within a fixed post-stimulus analysis window.
Each trial is assigned a binary supervision label
\begin{equation}
    y_i \in \{0,1\},
\end{equation}
where $y_i=1$ indicates matched evaluation and $y_i=0$ indicates mismatched evaluation.

Given labeled trials $\mathcal{D}=\{(x_i,y_i)\}_{i=1}^{N_{\text{trials}}}$, the decoding objective is to learn a subject-specific estimator
\begin{equation}
    f_\theta : \mathbb{R}^{C \times T} \rightarrow [0,1],
\end{equation}
parameterized by $\theta$, that maps each trial $x_i$ to a scalar output
\begin{equation}
    \hat{p}_i = f_\theta(x_i),
\end{equation}
which is interpreted as the estimated posterior probability of intention relevance, i.e., $\hat{p}_i \approx P(y_i=1\,|\,x_i)$.

Binary intention decisions are obtained via thresholding,
\begin{equation}
    \hat{y}_i = \mathbb{I}\!\left[\hat{p}_i \ge \tau\right],
\end{equation}
where $\mathbb{I}[\cdot]$ is the indicator function and $\tau=0.5$ unless otherwise specified.

Model parameters are learned by minimizing an empirical risk over $\mathcal{D}$,
\begin{equation}
    \min_{\theta}\; \frac{1}{N_{\text{trials}}}\sum_{i=1}^{N_{\text{trials}}}\mathcal{L}\!\left(y_i,\hat{p}_i\right),
\end{equation}
where $\mathcal{L}$ denotes a designed loss for model training.

\begin{figure*}[h!]
    \centering
    \includegraphics[width=1\linewidth]{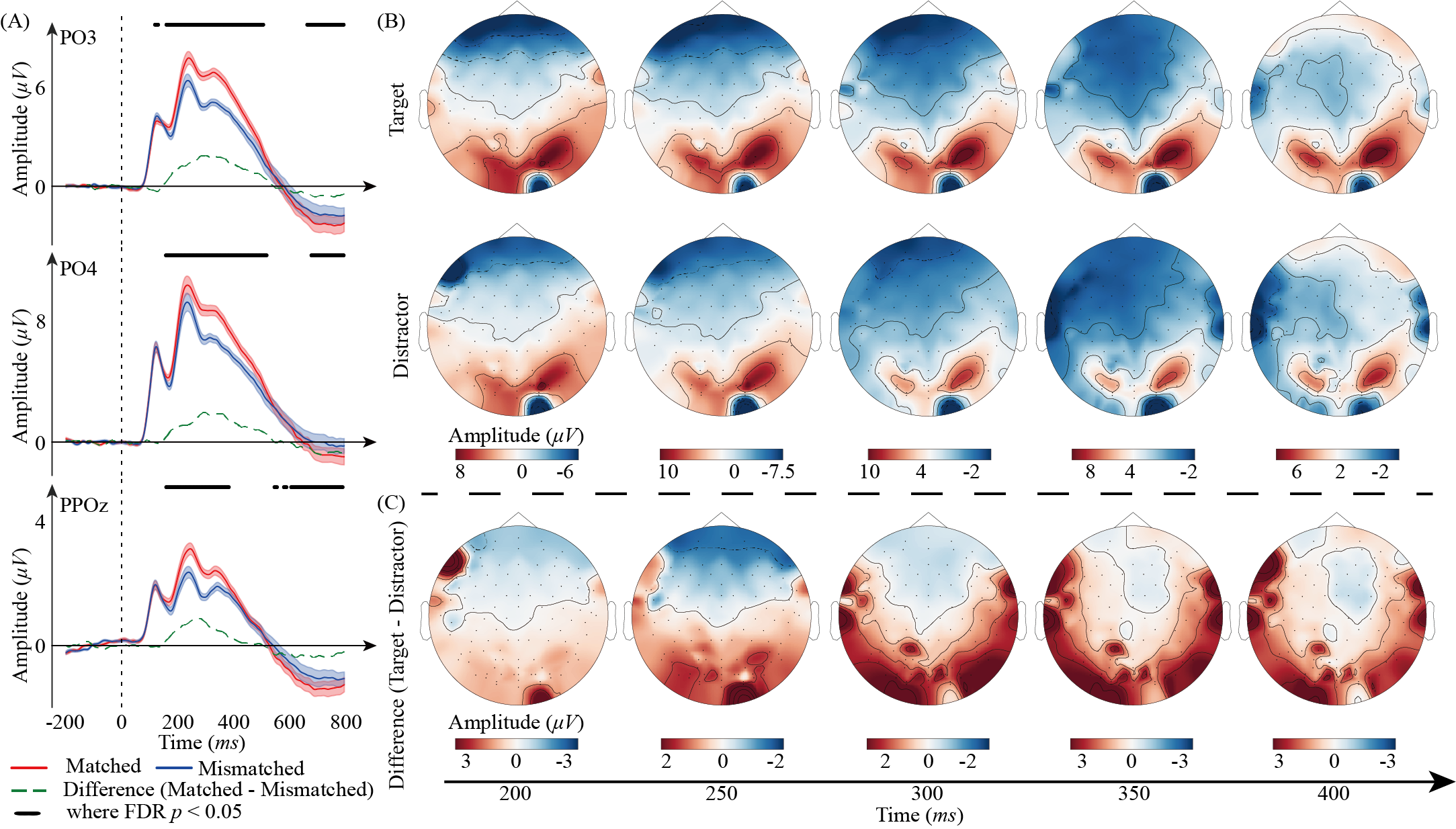}
    \caption{
    \textbf{Posterior P3b modulation by intention relevance.}
    (A) Grand-average ERPs at representative midline parietal-occipital electrodes (POz, CPPz, PPOz) show a pronounced P3b enhancement for intention-relevant (matched) relative to intention-irrelevant (mismatched) objects.
    (B) Scalp topographies depict the spatiotemporal evolution of ERPs for matched and mismatched conditions.
    (C) Matched-mismatched difference maps reveal a sustained posterior positivity in the canonical P3b time window.
    }
    \label{fig:p3b}
\end{figure*}

\subsection{Category-Cued Human Intention Paradigm}
\label{subsec:paradigm}

\subsubsection{Image Dataset}

The object-intention decoding task employed visual stimuli drawn from the Caltech-256 dataset \cite{Griffin2007Caltech256}, following~\cite{leong2023ventral,leong2024distinction}. 
Four semantically distinct object categories were selected—\textit{animals}, \textit{flowers}, \textit{food}, and \textit{vehicles}. 
Within each category, five representative object classes were chosen, and ten unique images were sampled per class. 

This resulted in a balanced stimulus set of 200 images ($4~categories \times 5~objects \times 10~images$), ensuring controlled variability in low-level visual features while preserving meaningful semantic structure for intention-based evaluation.

\subsubsection{Trial Structure}

At the onset of each trial block, a target category cue is presented centrally at fixation. 
Following this cue, a stream of object images is displayed sequentially, with each stimulus appearing at a randomized peripheral location around the fixation point.

Both object identity and spatial position are independently randomized across trials to prevent systematic visual or spatial prediction.
The experimental design is balanced such that the number of trials is held equal between the two intention conditions (matched and mismatched), ensuring unbiased estimation of decoding performance.

Crucially, target relevance is defined dynamically by the current category cue rather than by invariant stimulus properties.
As a result, multiple objects within the same sequence may be matched, such as the demo shown in Fig.~\ref{fig:model}(C), where the first three objects in the sequence all satisfy the matched condition.

EEG signals are time-locked to the onset of each object presentation, and intention is evaluated at the level of individual stimuli rather than at the sequence level.
This design enables single-trial inference of intention-related neural responses independent of sequence structure.
Detailed stimulus timing and spatial configuration are illustrated in Fig.~\ref{fig:model_exp_design}.

\subsubsection{Task Instructions}

Participants were instructed to maintain fixation and to internally evaluate whether each presented object was goal-relevant under the current task context.
All evaluations were performed covertly without overt motor responses, and participants were instructed to minimize eye movements and muscle activity throughout the task.

\subsection{Participants}
\label{subsec:participants}

Ninety-seven healthy adults participated in the offline experiment and Nine participants completed the online experiment. Six individuals took part in both the offline and online experiments.
All participants reported normal or corrected-to-normal vision and no history of neurological or psychiatric disorders.
Written informed consent was obtained prior to participation, and participants received monetary compensation.
All experimental procedures were approved by the University of Technology Sydney Human Research Ethics Committee (UTS HREC REF NO. ETH25-11711).

\subsection{EEG Signal Preprocessing}

EEG data were acquired at a sampling rate of 1000~Hz using a 128-channel Ag/AgCl electrode system (Neuroscan, Curry9) configured in the international 10-5 montage. 
Electrode impedances were maintained below 5~k$\Omega$ throughout the recording session to ensure stable signal quality.

The continuous EEG signals were digitally band-pass filtered between 0.5 and 40~Hz using a zero-phase finite-impulse-response (FIR) filter to remove slow drifts and high-frequency noise.
Filtered data were then segmented into stimulus-locked epochs spanning $-200$ to 800~ms relative to stimulus onset.
Each epoch was baseline-corrected using the pre-stimulus interval ($-200$ to 0~ms) and subsequently re-referenced to the common average across all electrodes.

\subsection{Neuroscientific Analysis}

Offline analyses were conducted on preprocessed EEG after additional artifact rejection.
Ocular components were removed using independent component analysis (ICA), and contaminated trials were excluded based on amplitude and variance thresholds ($\pm 100~\mu\mathrm{V}$).
Trials exhibiting residual muscle artifacts, abrupt signal discontinuities, or abnormal channel-wise variance were discarded.

ERPs were obtained by averaging epochs time-locked to stimulus onset for matched and mismatched trials.
To isolate late evaluative activity associated with goal-dependent processing, mean ERP amplitudes were quantified at electrodes PO3, PO4, and PPOz within the 300--600~ms post-stimulus window~\cite{Zhang2015NeuroImage,Polich2007UpdatingP300}.

To examine the spatiotemporal characteristics of intention-related modulation, condition-specific scalp topographies and matched-minus-mismatched difference maps were computed for representative late windows (200 \textasciitilde 400~ms and 300 \textasciitilde 600~ms).
Statistical differences between conditions were assessed using paired $t$-tests at each electrode and time window, with false discovery rate (FDR) correction applied for multiple comparisons.

\subsection{Model Structure}
\label{subsec:model}

The proposed framework (Fuzzy Prototype-Based Latent Intention, Fuzz-PLI) is formulated as an intention decoding system for estimating goal-relevance from high-dimensional EEG observations.
It adopts a modular architecture consisting of a Convolutional feature encoder, an interpretable fuzzy prototype-based inference module, and a lightweight classification head.
As illustrated in Fig.~\ref{fig:model}(A), the same architecture is used for both offline analysis and online validation, enabling real-time intention inference without architectural modification.

\subsubsection{Latent Feature Extraction}
\label{subsec:encoder}

Each preprocessed EEG epoch $x_i \in \mathbb{R}^{C \times T}$ is first mapped into a structured latent representation that serves as the antecedent space for subsequent fuzzy inference.
Rather than performing fuzzy reasoning directly in the raw signal domain, we construct a compact and task-relevant latent feature space in which fuzzy rules can operate more effectively.

We instantiate the encoder $p_0(\cdot)$ using an EEGNet-inspired convolutional architecture~\cite{lawhernEEGNetCompactConvolutional2018}, consisting of temporal convolution, depthwise spatial convolution, and separable convolution blocks.
Temporal convolution captures band-limited neural dynamics, depthwise spatial convolution learns subject-specific spatial filters across electrodes, and separable convolution integrates temporal and spatial information into a structured representation.

Formally, each input epoch is mapped to
\begin{equation}
\mathbf{X}_i = p_0(x_i) \in \mathbb{R}^{F \times H \times W},
\end{equation}
where the resulting latent units are retained as a structured tensor rather than being flattened.
In the Fuzz-PLI framework, each latent unit is interpreted as an antecedent variable in a high-dimensional fuzzy inference system.
All encoder parameters are shared across trials within a subject and optimized jointly with the fuzzy inference module.

\begin{figure}[t!]
    \centering
    \includegraphics[width=1\linewidth]{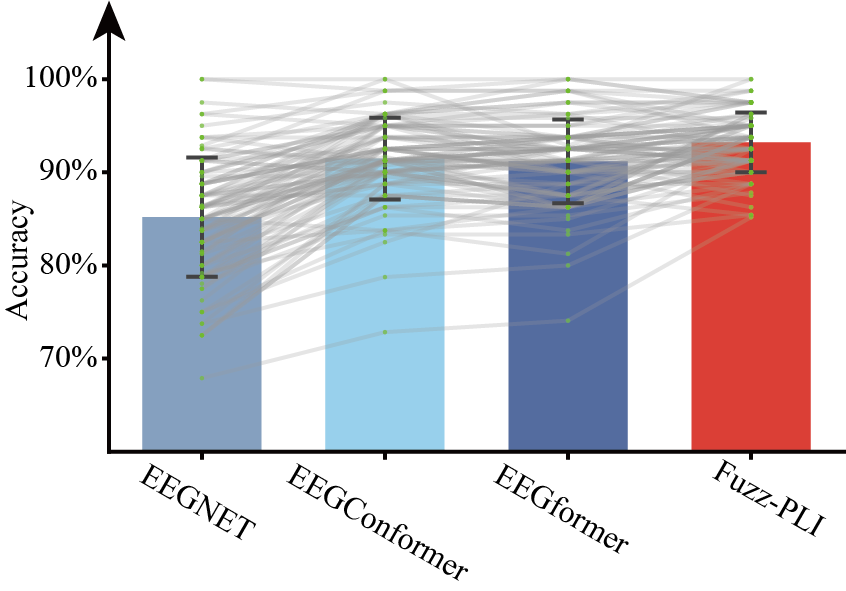}
    \caption{
        \textbf{Offline within-subject decoding accuracy across models.}
        Bars denote mean classification accuracy across participants, with error bars indicating one standard deviation of the mean.
        Green dots represent individual participants, and gray lines connect paired results for the same participant across models.
        }
    \label{fig:model_comparison}
\end{figure}

\subsubsection{Fuzzy Prototype Modulation in Fuzzy Prototype-Based Latent Intention (Fuzz-PLI)}
\label{subsec:fuzzy}

Building upon the Fuzzy Inference Systems (FISs) framework~\cite{shihabudheen2018recent, 10.1007/978-3-642-58930-0_10, Chaudhari2014StudyAR,928739,704563}, we introduce a generalized fuzzy prototype modulation module operating in the latent feature space.
Rather than mapping inputs directly to scalar rule consequents, the proposed formulation realizes fuzzy inference as a rule-conditioned modulation of latent representations, enabling graded and interpretable intention reasoning.

\paragraph{Latent Antecedent Space}
The structured latent feature map $\mathbf{X}_i = p_0(x_i)$ is interpreted as a high-dimensional antecedent space.
Each latent unit corresponds to an antecedent variable in a generalized TSK system, allowing fuzzy rules to operate over distributed neural representations instead of handcrafted features.

\paragraph{Rule Parameterization in the Raw EEG Domain}
We maintain $R$ fuzzy rules.
Each rule $r$ is parameterized by a learnable prototype center
$\tilde{\mathbf{c}}_r \in \mathbb{R}^{1 \times C \times T}$
and a learnable width tensor
$\tilde{\mathbf{w}}_r \in \mathbb{R}^{1 \times C \times T}$,
both defined in the raw EEG domain.
To ensure comparability between trial features and rule prototypes, both parameters are passed through the shared convolutional encoder $p_0(\cdot)$,
\begin{equation}
\mathbf{C}_r = p_0(\tilde{\mathbf{c}}_r), \qquad
\boldsymbol{\Sigma}_r = \mathrm{softplus}\!\left(p_0(\tilde{\mathbf{w}}_r)\right) + \sigma_{\min},
\end{equation}
where the softplus transformation enforces strictly positive widths and $\sigma_{\min}$ prevents degenerate solutions.

\paragraph{Antecedent Membership Computation}
For each trial $i$ and rule $r$, antecedent memberships are computed via a Gaussian function in the latent space,
\begin{equation}
\boldsymbol{\mu}_{r,i}
=
\exp\!\left(
-\frac{(\mathbf{X}_i - \mathbf{C}_r)^2}{2\boldsymbol{\Sigma}_r^2 + \varepsilon}
\right),
\end{equation}
where each latent unit contributes one antecedent membership degree.
This formulation quantifies the compatibility between trial-specific latent patterns and rule prototypes.

\paragraph{Rule Firing Strength}
Let $\{\mu_{r,i}^{(d)}\}_{d=1}^{D}$ denote the set of antecedent memberships associated with rule $r$, where $D$ is the total number of latent antecedents.
The firing strength of rule $r$ for trial $i$ is obtained using a mean t-norm,
\begin{equation}
s_{r,i}
=
\frac{1}{D}\sum_{d=1}^{D}\mu_{r,i}^{(d)},
\end{equation}
which provides a smooth and numerically stable alternative to product-based aggregation in classical TSK systems.
Firing strengths are calculated across rules via
\begin{equation}
\alpha_{r,i}
=
\frac{\exp(s_{r,i}/\tau)}{\sum_{k=1}^{R}\exp(s_{k,i}/\tau)},
\end{equation}
where $\tau$ controls the sharpness of rule competition.

\paragraph{Fuzzy Consequent and Modulation}
Unlike conventional TSK systems with scalar consequents, each rule in the Fuzz-PLI contributes a vector-valued consequent that modulates the latent representation.
The aggregated fuzzy activation is computed as
\begin{equation}
\boldsymbol{\mu}^{\mathrm{comb}}_i
=
\sum_{r=1}^{R}\alpha_{r,i}\,\boldsymbol{\mu}_{r,i},
\end{equation}
and the final rule-conditioned consequent is applied via feature-wise modulation,
\begin{equation}
\mathbf{X}_i^{\mathrm{mod}}
=
\mathbf{X}_i
+
\lambda\,\mathbf{X}_i \odot \boldsymbol{\mu}^{\mathrm{comb}}_i,
\end{equation}
where $\lambda$ controls the strength of fuzzy modulation.
This formulation preserves the original latent structure while selectively amplifying intention-consistent neural patterns.

\paragraph{Training Strategy and Prototype Diversity Regularization}
To improve optimization stability, fuzzy modulation is activated only after a warm-up stage, during which the convolutional encoder and classification head are trained without fuzzy inference.
This staged training strategy prevents early domination of unstable fuzzy rules and facilitates robust latent feature learning.

To prevent rule collapse and encourage diversity among fuzzy prototypes, an inter-prototype regularization term is imposed.
Specifically, each raw prototype center $\tilde{\mathbf{c}}_r$ is first vectorized and $\ell_2$-normalized,
\begin{equation}
\hat{\mathbf{c}}_r =
\frac{\mathrm{vec}(\tilde{\mathbf{c}}_r)}{\|\mathrm{vec}(\tilde{\mathbf{c}}_r)\|_2},
\end{equation}
and the average pairwise cosine similarity between normalized prototypes is penalized,
\begin{equation}
\mathcal{L}_{\mathrm{div}}
=
\frac{1}{R(R-1)}
\sum_{r \neq k}
\hat{\mathbf{c}}_r^{\top} \hat{\mathbf{c}}_k .
\end{equation}
This regularization encourages distinct fuzzy rules to capture complementary latent intention patterns.

\begin{figure*}[ht!]
    \centering
    \includegraphics[width=0.85\linewidth]{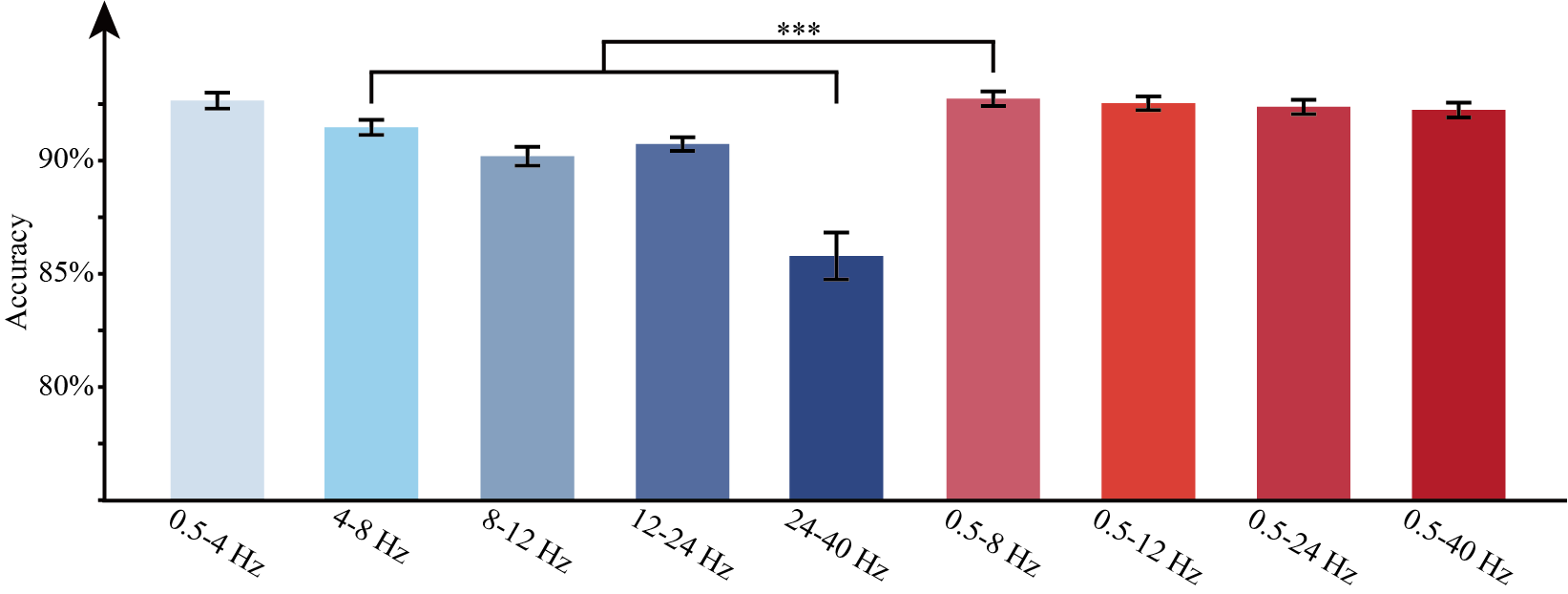}
    \caption{
        Effect of frequency band selection on within-subject decoding accuracy.
        Each point corresponds to an individual participant. Error bars denote the standard error of the mean (SEM).
        Low-frequency bands yield consistently higher performance, whereas higher-frequency bands show marked performance degradation.
    }
        
    \label{fig:ab_freq}
\end{figure*}

\subsubsection{Training and Evaluation Protocol}
\label{subsec:training}

All experiments were conducted in a within-subject setting.
For each participant, trials ($N=400$) were randomly partitioned into training (80\%) and validation (20\%) sets with balanced class labels. This procedure was repeated five times using different random seeds, and the final performance was reported as the average across the five runs.
The encoder, fuzzy prototype parameters, and classification head were optimized jointly using binary cross-entropy loss,
\begin{equation}
\mathcal{L}_{\mathrm{CE}} =
- \frac{1}{|\mathcal{D}_{\mathrm{train}}|}
\sum_{i \in \mathcal{D}_{\mathrm{train}}}
\left[
y_i \log \hat{p}_i + (1 - y_i)\log(1 - \hat{p}_i)
\right].
\end{equation}
where $\mathcal{D}_{\mathrm{train}}$ denote the set of training trials obtained from the within-subject data split.
The final objective combines classification loss and prototype diversity regularization,
\begin{equation}
\mathcal{L}
=
\mathcal{L}_{\mathrm{CE}}
+
\lambda_{\mathrm{div}}\,\mathcal{L}_{\mathrm{div}},
\end{equation}
with $\lambda_{\mathrm{div}} = 0.3$.
Performance was evaluated using within-subject classification accuracy, and group-level statistics were computed across participants.

\subsection{Baseline Models}

The Fuzz-PLI model was compared with three representative EEG decoding baselines: EEGNet~\cite{lawhernEEGNetCompactConvolutional2018}, EEG Conformer~\cite{song2023eeg}, and EEGformer~\cite{Wan2023EEGformer}.
EEGNet employs depthwise and separable convolutions for compact spatiotemporal modeling.
EEG Conformer ($N_{\mathrm{Deep}}=6$, $D_{\mathrm{emb}}=256$) integrates convolutional front-ends with self-attention mechanisms to capture long-range dependencies, while EEGformer ($N_{\mathrm{block}}=3$, $N_{\mathrm{head}}^{\mathrm{RTM}}=4$, $N_{\mathrm{head}}^{\mathrm{STM}}=4$, $N_{\mathrm{head}}^{\mathrm{TTM}}=11$) adopts a hierarchical transformer architecture for regional and temporal feature modeling.

All baseline models were implemented following their original designs.
Only dataset-dependent input dimensions were adjusted.
To ensure a fair comparison, all models shared identical EEG preprocessing pipelines, including channel configuration, band-pass filtering, baseline removing, and temporal window selection.
Training configurations were strictly matched across models, with the same learning rate (0.00015), batch size (16), number of epochs (120), and optimization strategy (AdamW~\cite{Loshchilov2017DecoupledWD}).

\section{Results}

\subsection{P3b Responses to Matched and Mismatched Object}

Target stimuli elicited a robust posterior P3b response, whereas distractor stimuli evoked substantially weaker activity. As shown in Fig.~\ref{fig:p3b}(A), event-related potentials over midline parietal electrodes (POz, CPPz, PPOz) exhibited a sustained positive deflection beginning at approximately 250~ms post-stimulus, which remained significant after false discovery rate (FDR) correction ($p<0.05$). This modulation resulted in a significant Matched-Mismatched Object difference within the canonical P3b time window (300\textasciitilde600~ms), with a maximal paired-sample contrast of $t(96)=6.59$, $p<0.001$, Cohen’s $d=0.39$. An earlier window (280\textasciitilde380~ms) showed an even stronger divergence between conditions ($t(96)=7.85$, $p<0.001$, Cohen’s $d=0.51$), consistent with an early emergence of goal-relevant evaluative activity.

Hemispheric analyses further revealed pronounced bilateral posterior effects. Over left parietal electrodes (P3, PO3, CPP3h), target stimuli produced a markedly larger P3b response ($t(96)=13.01$, $p<0.001$, Cohen’s $d=0.67$). A comparably strong effect was observed over right parietal sites (P4, PO4, CPP4h; $t(96)=11.91$, $p<0.001$, Cohen’s $d=0.58$). The presence of robust effects in both hemispheres indicates that target evaluation engaged a distributed posterior decision-related network rather than a strictly lateralized process.

The accompanying scalp maps, as presented in Fig.~\ref{fig:p3b}(B), show that target trials elicited a widespread centro-parietal positivity, whereas distractor responses remained comparatively attenuated. Difference topographies (Fig.~\ref{fig:p3b}(C)) highlight a stable posterior cluster spanning Occipital to midline Parietal sites between 200 and 400~ms, consistent with the canonical spatial signature of the P3b component.

Together, these results demonstrate a reliable and spatially extended enhancement of late parietal activity when stimuli align with internally specified goals. From a cybernetic perspective, this posterior positivity serves as an observable neural manifestation of a latent intention-related evaluative state, rather than a purely stimulus-driven or sensory response. Importantly, because physical stimulation is held constant across intention conditions, the observed modulation reflects goal-dependent state inference rather than object-specific features.

\begin{figure}
    \centering
    \includegraphics[width=1\linewidth]{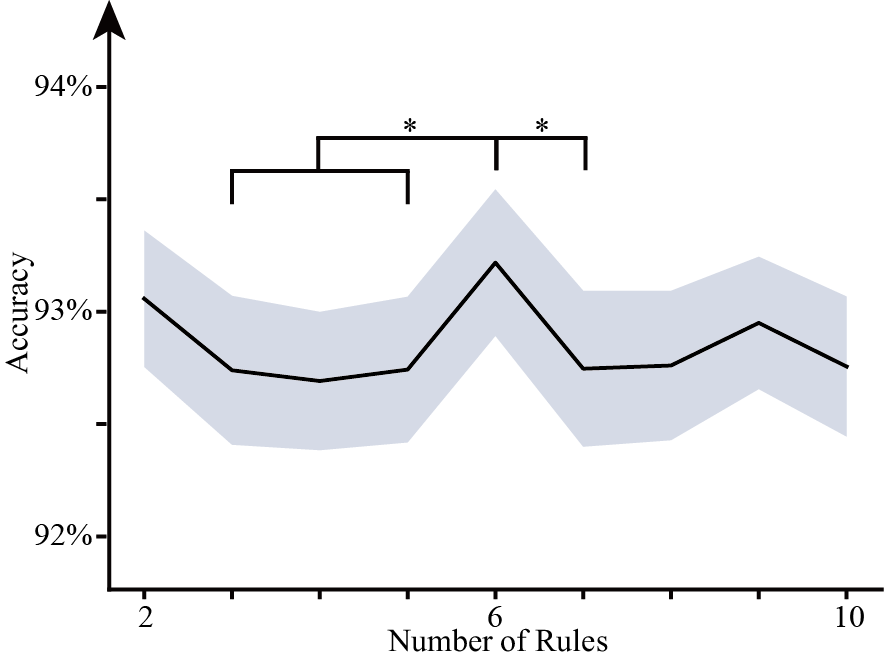}
    \caption{
    Effect of the number of fuzzy rules on within-subject decoding accuracy.
    Performance plateaus with a small number of rules, indicating that increasing rule count beyond this range yields no substantial benefit.
    }
    \label{fig:ab_rule}
\end{figure}

\subsection{Model Comparison}

The proposed Fuzz-PLI model achieved the highest within-subject decoding accuracy among all evaluated approaches ($N_{params}=1.46M$, 93.22\% $\pm$ 3.21\%; $n=97$; Fig.~\ref{fig:model_comparison}). Performance was consistently superior to EEGNet ($N_{params}=0.11M$,85.190\% $\pm$ 6.390\%), EEGConformer ($N_{params}=6.25M$,91.47\% $\pm$ 4.39\%), and EEGformer ($N_{params}=4.53M$, 91.17\% $\pm$ 4.49\%).

Paired $t$-tests confirmed that Fuzz-PLI significantly outperformed EEGNet ($t(96)=13.959$, $p<0.001$), EEGConformer ($t(96)=5.129$, $p<0.001$), and EEGformer ($t(96)=5.788$, $p<0.001$). Effect sizes ranged from moderate to large (Cohen’s $d_z=0.455$-$1.587$), indicating robust improvements across participants rather than performance gains driven by a small subset of subjects. All $p$ values are FDR-corrected.

These results indicate that explicitly modeling intention as a latent internal state with graded belief representations yields more robust single-trial inference than architectures relying solely on implicit feature learning.

% \begin{table}[t]
% \centering
% \caption{Computational cost comparison in terms of FLOPs for a single forward inference.}
% \label{tab:flops_only}
% \begin{tabular}{lcc}
% \hline
% Model & \#Rules & FLOPs (G) \\
% \hline
% EEGNet        & - & 0.233 \\
% EEGConformer & - & 9.062 \\
% EEGFormer    & - & 1.23 \\
% \hline
% Fuzzy-PLI    & 2  & 1.167 \\
% Fuzzy-PLI    & 3  & 1.633 \\
% Fuzzy-PLI    & 4  & 2.100 \\
% Fuzzy-PLI    & 5  & 2.566 \\
% Fuzzy-PLI    & 6  & 3.033 \\
% Fuzzy-PLI    & 7  & 3.500 \\
% Fuzzy-PLI    & 8  & 3.966 \\
% Fuzzy-PLI    & 9  & 4.433 \\
% Fuzzy-PLI    & 10 & 4.899 \\
% \hline
% \end{tabular}
% \end{table}

\subsection{Ablation Study}

We conducted a series of ablation experiments to evaluate the sensitivity of Fuzz-PLI to key architectural and signal-processing choices.

\paragraph{Frequency Band}
Decoding performance was highest in low-frequency bands, especially at Delta (0.5\textasciitilde4~Hz) and Theta (0.5\textasciitilde8~Hz) bands. 0.5\textasciitilde8~Hz filtering yielded the strongest performance (93.22\% $\pm$ 3.21\%).
Performance differences among broader frequency ranges (0.5\textasciitilde4~Hz, 0.5\textasciitilde12~Hz, 0.5\textasciitilde24~Hz, 0.5\textasciitilde40~Hz) did not reach statistical significance (all $p>0.05$). In contrast, higher-frequency bands (Alpha band at 8\textasciitilde12~Hz, Beta band at 12\textasciitilde24~Hz, Gamma at 24\textasciitilde40~Hz) resulted in marked and significant performance degradation (all $p<0.001$ after FDR-corrected; Fig.~\ref{fig:ab_freq}).

These results suggest that Fuzz-PLI primarily exploits low-frequency neural dynamics that are characteristic of sustained latent evaluative states, rather than transient stimulus-driven activity.

\paragraph{Number of Fuzzy Rules}
Varying the number of fuzzy rules from 2 to 10 had minimal impact on decoding accuracy. Performance plateaued for the number of rules $\geq 2$, with no systematic or practically meaningful improvement observed beyond this range (all $\Delta Accuracy < 0.53\%$; Fig.~\ref{fig:ab_rule}). Decoding performance peaked at six fuzzy rules. This insensitivity indicates that a small number of fuzzy rules is sufficient to capture the dominant latent structure of within-subject neural representations, supporting model parsimony and interpretability. This property is consistent with a latent-state inference perspective, in which a small number of prototypes is sufficient to represent the dominant internal states governing intention-related evaluation.

\paragraph{Firing Strength Strategy}
We compared TSK-inspired softmax-weighted strategy with the original Mamdani-style max-based strategy~\cite{jiang2025ifuzzmeta} under identical settings (six rules; 0.5\textasciitilde8~Hz). Softmax aggregation substantially outperformed max-based aggregation (93.22\% $\pm$ 3.21\% vs.\ 88.84\% $\pm$ 5.42\%; $t(96)=9.516$, $p<0.001$), demonstrating that preserving graded rule contributions is critical for effective fuzzy modeling.

\paragraph{Effect of the Fuzzy Prototype Module}
Finally, removing the fuzzy prototype module while keeping the backbone architecture fixed led to a consistent reduction in within-subject decoding accuracy (93.22\% $\pm$ 3.21\% vs.\ 88.84\% $\pm$ 5.42\%; $t(96)=9.516$, $p<0.001$, Cohen’s $d=0.983$). This confirms that the observed performance gains arise from the proposed fuzzy mechanism itself rather than increased architectural capacity alone.

Together, these ablation results indicate that Fuzz-PLI achieves robust within-subject decoding through low-frequency neural modeling, soft rule aggregation, and a parsimonious fuzzy structure, rather than reliance on fine-grained architectural complexity.

\section{Online Validation}
\label{sec:online}
\subsection{Experimental Setup}

To assess real-time feasibility, we conducted an online validation experiment using the same task paradigm as in the offline analysis. Each trial consisted of a 1~s decoding window, preceded by a variable fixation baseline (0.5\textasciitilde1~s). Following the results of offline evaluation, we use 0.5\textasciitilde8 Hz band-pass filter.

Nine participants completed the online experiment. For each participant, two online sessions were collected, with each session comprising 400 trials (200 target and 200 non-target trials). The model was trained in a within-subject manner using data from the first session. Immediately after training, the trained model was deployed online and evaluated on data from the second session without any further parameter tuning or recalibration. In the online experiment, both sessions followed the same task paradigm and stimulus set, emphasizing the model’s real-time decoding stability and session-to-session transfer within each participant.

During online operation, real-time feedback was generated exclusively using the proposed Fuzz-PLI framework. Baseline models were not deployed for real-time feedback; instead, their performance was evaluated offline using data collected during the online sessions and processed with the same pipeline to ensure a fair comparison with paired $t$-test.

\subsection{Online System}

The online validation system is illustrated in Fig.~\ref{fig:model}(B).
During online inference, EEG signals were continuously acquired into buffer (20s) and captured as short epochs (-0.2s\textasciitilde0.8s) by trigger signal, which were then processed in real time by the trained Fuzz-PLI model.

\subsection{Results}

\begin{figure}[t]
    \centering
    \includegraphics[width=1\linewidth]{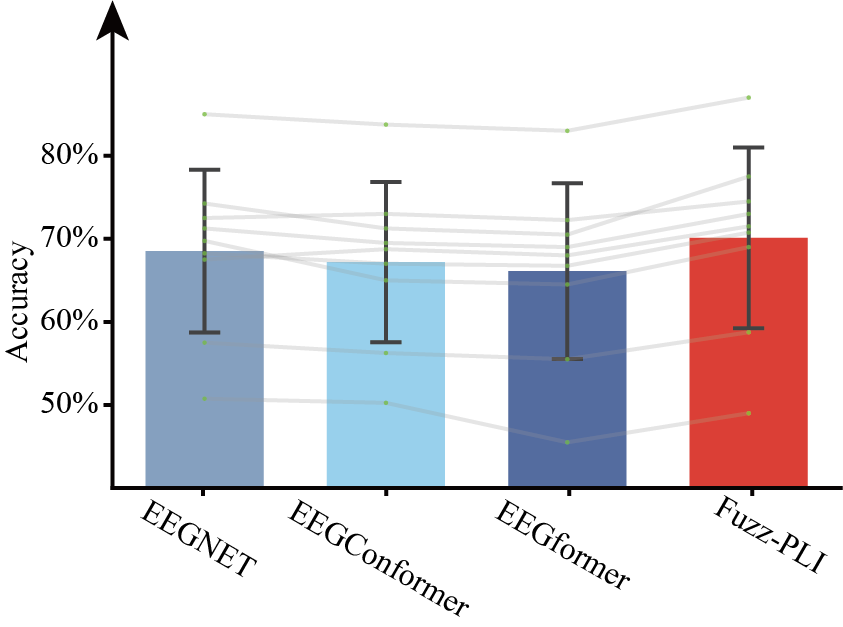}
    \caption{Online decoding accuracy across models.
    Bars denote mean classification accuracy across participants, with error bars indicating one standard deviation.
    Green dots represent individual participants, and gray lines connect paired results from the same participant across models.}
    \label{fig:online_res}
\end{figure}

\paragraph{Online decoding performance}
Across participants, Fuzz-PLI achieved an average online decoding accuracy of
$70.11\% \pm 10.87\%$, consistently outperforming all baseline models, as shown in Fig.~\ref{fig:online_res}.
Among the baseline methods, EEGNet yielded the second-highest accuracy ($68.53\% \pm 9.80\%$), but its performance was significantly lower than that of Fuzz-PLI ($t(8)=2.610$, $p=0.031$, Cohen’s $d=0.153$).
Fuzz-PLI also significantly outperformed EEGConformer ($67.19\% \pm 9.64\%$; $t(8)=4.308$, $p=0.003$, Cohen’s $d=0.284$) and EEGFormer ($66.11\% \pm 10.57\%$; $t(8)=9.281$, $p<0.001$, Cohen’s $d=0.373$), indicating robust advantages across both convolutional and transformer-based architectures.

\paragraph{Comparison with Mamdani-style Strategy}
To further examine the contribution of the TSK-style strategy, we compared it with the original Mamdani-style approach~\cite{jiang2025ifuzzmeta} implemented under identical experimental settings.
TSK-style strategy achieved significantly higher online decoding accuracy than the max-based strategy ($69.17\% \pm 10.19\%$; $t(8)=2.538$, $p=0.035$, Cohen’s $d=0.090$), suggesting that soft aggregation provides measurable performance benefits over conventional max-based fuzzy inference in real-time decoding scenarios.

\begin{figure*}[ht!]
    \centering 
    \includegraphics[width=0.85\linewidth]{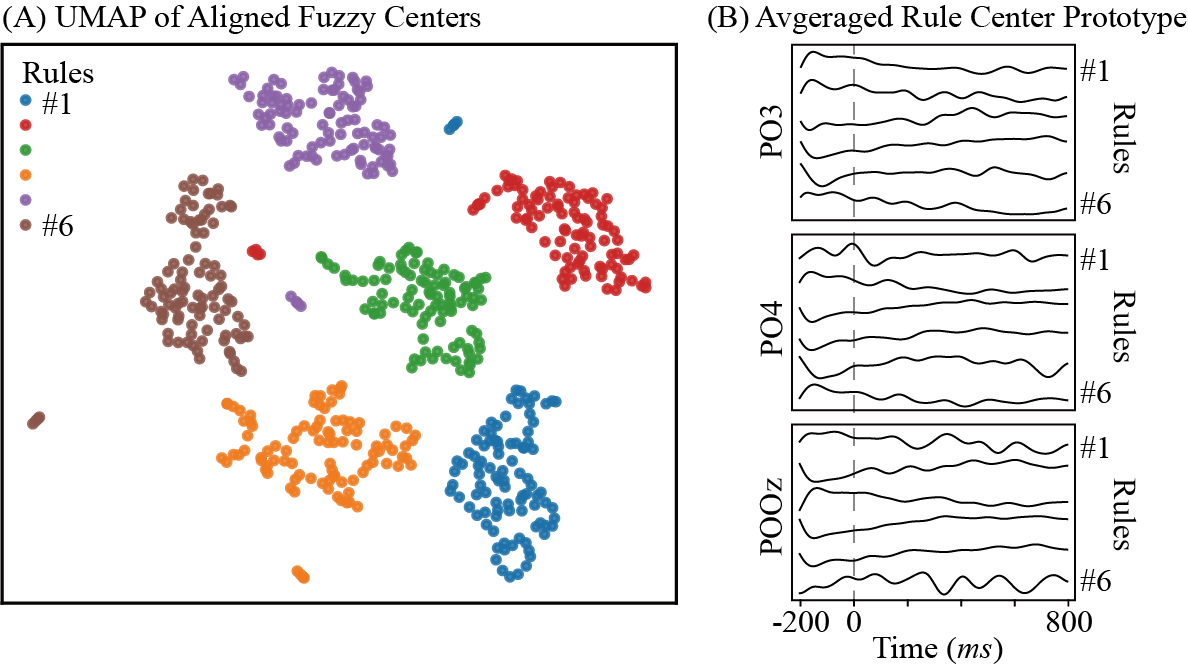} 
    \caption{
    Interpretability analysis of the fuzzy rule representations.
    (A) UMAP embedding of aligned fuzzy rule prototypes across participants, with each dot representing one participant.
    Rule identities were aligned using cosine similarity, revealing cross-subject consistency.
    (B) Averaged fuzzy rule prototypes at electrodes PO3, PO4, and POOz, illustrating distinct spatiotemporal patterns across rules.
    }
    
    \label{fig:interp_group} 
\end{figure*}

Taken together, these factors reflect fundamental challenges in closed-loop cybernetic brain-computer interfaces rather than deficiencies of any specific decoding architecture.
Accordingly, performance degradation under online conditions should be interpreted as a manifestation of real-time latent state inference under distribution shift and uncertainty, which is intrinsic to human-machine interactive systems.

\section{Discussion}

\subsection{P3b as an Observable of a Latent Intention-Related State}

The P3b component has been extensively characterized as a neural signature of goal-relevant, post-perceptual evaluation rather than a purely stimulus-driven sensory response.
In target detection paradigms, P3b amplitude scales with task relevance even when physical stimulus properties are held constant, consistent with accounts linking the P3b to stimulus-response (S-R) link reactivation and internal decision model updating~\cite{Verleger2005,Verleger2020}.
From the perspective adopted here, this late parietal activity should not be interpreted as intention itself, but rather as an observable neural manifestation of an underlying goal-dependent evaluative state that determines whether sensory evidence is committed to action-relevant processing.

Evidence from no-report and dissociation paradigms further indicates that the P3b does not constitute a neural correlate of perceptual awareness or intention formation, but instead reflects post-perceptual evaluation and decision-related processing~\cite{Cohen2020}.
Accordingly, the P3b indexes intention \emph{realization} rather than intention initiation, marking the point at which accumulated sensory evidence is evaluated with respect to an internal goal.
This distinction is critical for BCI design: rather than decoding awareness or stimulus detection, the P3b provides a temporally precise and physiologically grounded observation channel for inferring latent intention-related states under time constraints.

\subsection{Interpretability of the Fuzzy Prototype-Based Model}

The interpretability analysis provides converging evidence that Fuzz-PLI model captures structured and non-random representations of human intention.
As shown in Fig.~\ref{fig:interp_group}(A), although model training was performed in a within-session manner, fuzzy rule prototypes exhibit substantial cross-subject consistency after alignment.
This finding suggests that the model converges onto shared latent neural patterns associated with evaluative processing, rather than overfitting idiosyncratic noise or session-specific artifacts.

At the rule level, Fig.~\ref{fig:interp_group}(B) illustrates that different fuzzy prototypes encode distinct spatiotemporal neural patterns.
For example, one rule exhibits a pronounced parieto-occipital response over electrode POOz in the 200-400~ms post-stimulus interval, closely resembling the canonical P3b component identified in the neuro-scientific analysis.
Other rules capture complementary patterns with distinct temporal profiles and spatial distributions, indicating that the model decomposes high-dimensional neural activity into multiple functionally meaningful components.

Together, these observations indicate that the fuzzy module does not rely on a single dominant neural signature, but instead performs a structured partitioning of evaluative neural dynamics into a compact set of interpretable prototypes.
Such rule-level decomposition provides a principled link between decoding performance and neuroscientific interpretation, enabling transparent inspection of how latent intention-related states are represented and inferred from neural data.

\subsection{System-Level Interpretation and Cybernetic Implications}

The findings support a cybernetic view in which intention is a latent internal state, with posterior P3b serving as its neural observation channel.
Because stimulus properties are decorrelated from relevance, the observed effects reflect goal-dependent state transitions rather than sensory processing.
Within this framework, Fuzz-PLI performs intention inference from noisy EEG.
A small set of fuzzy prototypes captures low-dimensional intention dynamics, explaining why performance saturates with only a few rules.
The model’s advantage over deep baselines indicates that explicit latent-state modeling yields more robust intention decoding.
Online validation further demonstrates that this framework is deployable in real-time, enabling practical intention-centric BCI control.

\subsection{Online Validation}

\subsubsection{Online versus Offline Performance}
Online decoding accuracy was lower than offline performance, which is a common phenomenon in real-time BCI systems~\cite{11205327,cao2025novel,Ding2025NatCommunBCI}. Offline evaluation benefits from non-causal preprocessing and extensive artifact removal, whereas online inference must satisfy strict latency and causal constraints, leading to a reduced effective signal-to-noise ratio. Despite these limitations, the proposed Fuzz-PLI method maintained superior performance compared to all baseline approaches, demonstrating its robustness for real-time neural intention decoding.

\subsubsection{System-level Sources of Online Performance Degradation}
The online-offline performance gap can be mainly attributed to three factors. 
First, EEG signals are inherently non-stationary across sessions, driven by changes in participant state (e.g., vigilance, fatigue, and cognitive strategy), a well-known challenge in high-level cognitive decoding. 
Second, the online pipeline introduces unavoidable system latency and imperfect temporal locking between stimulus/event markers and neural recordings, leading to temporal jitter in the effective decoding window. 
Third, despite regularization, the offline-trained model may still exhibit inevitable overfitting to session-specific patterns, reducing generalization under deployment conditions. 
Collectively, these factors account for the observed degradation in online performance.

\subsection{Limitations and Future Directions}

This study is limited to within-subject intention inference; extending the framework to cross-subject adaptation and long-term deployment remains for future work.
The current paradigm also treats intention as an instantaneous state, whereas future studies should model intention as a continuous latent process.
In addition, the number of fuzzy rules was fixed, and adaptive or hierarchical rule learning may further improve generalization. 
Future research will address the gap between offline and online performance.
At the system level, latency-aware inference, lightweight online calibration, and drift-aware thresholds are expected to improve robustness.
At the user level, neurofeedback training may help stabilize intention-related neural signatures.
These directions will support more reliable and scalable intention-centric BCI systems.

{\small
\bibliographystyle{IEEEtran}
\bibliography{ref}
}

\end{document}